\documentclass[sn-mathphys-num]{sn-jnl}

\usepackage{graphicx}%
\usepackage{multirow}%
\usepackage{amsmath,amssymb,amsfonts}%
\usepackage{amsthm}%
\usepackage{mathrsfs}%
\usepackage[title]{appendix}%
\usepackage{xcolor}%
\usepackage{textcomp}%
\usepackage{manyfoot}%
\usepackage{booktabs}%
\usepackage{listings}%
\usepackage{subcaption}
\usepackage{geometry}
\usepackage{pdflscape}

\usepackage[ruled,linesnumbered]{algorithm2e} 

\SetAlFnt{\small}
\SetAlCapFnt{\small}
\SetAlCapNameFnt{\small}
\SetAlCapHSkip{0pt}
\IncMargin{-\parindent}

\theoremstyle{thmstyleone}%
\theoremstyle{thmstyletwo}%

\theoremstyle{thmstylethree}%

\begin{document}

\title[Improving ZNE for Quantum-gate Error Mitigation using a Noise-aware Folding Method]{Improving Zero-noise Extrapolation for Quantum-gate Error Mitigation using a Noise-aware Folding Method}


\author[1]{\fnm{Leanghok} \sur{Hour}}


\author[1]{\fnm{Myeongseong} \sur{Go}}

\author*[1]{\fnm{Youngsun} \sur{Han}}\email{youngsun@pknu.ac.kr}


\affil[1]{\orgdiv{Department of AI Convergence}, \orgname{Pukyong National University}, \orgaddress{\city{Busan}, \country{South Korea}}}


\abstract{
Recent thousand-qubit processors represent a significant hardware advancement, but current limitations prevent effective quantum error correction (QEC), necessitating reliance on quantum error mitigation (QEM) to enhance result fidelity from quantum computers.
Our paper introduces a noise-aware folding technique that enhances Zero-Noise Extrapolation (ZNE) by leveraging the noise characteristics of target quantum hardware to fold circuits more efficiently.
Unlike traditional ZNE approaches assuming uniform error distribution, our method redistributes noise using calibration data based on hardware noise models.
By employing a noise-adaptive compilation method combined with our proposed folding mechanism, we enhance the ZNE accuracy of quantum gate-based computing using superconducting quantum computers.
This paper highlights the uniqueness of our method, summarizes noise accumulation, presents the scaling algorithm, and compares the reliability of our method with those of existing models using linear extrapolation model. 
Experimental results show that compared to existing folding methods, our approach achieved a 35\% improvement on quantum computer simulators and a 31\% improvement on real quantum computers, demonstrating the effectiveness of our proposed approach.}

\keywords{Quantum computing, quantum error mitigation, zero-noise extrapolation, target-aware optimization}



\maketitle

\section{Introduction}\label{sec1:intro}

Recently, quantum computing (QC) technology has significant advancements, including IBM's release of processors housing over a thousand qubits~\cite{lau2022nc, huang2020sq}, as well as the development of more robust crosstalk chips utilizing tunable-coupler technology with up to 133 qubits~\cite{ezratty2023po, sussman2023qc,mckay2023}. 
Despite these advancements, the capacity of current hardware is insufficient for implementing QECs which is projected to become available by the 2030s~\cite{johansson2021qc, jyothi2022io}. 
In the meantime, QC systems employ QEM methods, often requiring additional quantum and classical resources as a trade-off to enhance QC output fidelity.

Numerous QEM techniques for reducing diverse sources of error in QC have been proposed~\cite{qin2022ao}: 
probabilistic error cancellation for mitigating decoherence ~\cite{van2023pe, temme2017em}, ZNE for imperfect gates, techniques for mitigating measurement errors ~\cite{ramadhani2021qe, tannu2019mm,sun2021mr}, dynamical decoupling~\cite{niu2022eo, qi2023eo}, quantum optimal control~\cite{hartnett2023is, dong2020rc}, randomized compiling~\cite{hashim2020rc, urbanek2021md}, Pauli-frame randomization~\cite{ware2021ep,suzuki2022qe}, and other techniques~\cite{endo2018pq, cai2023qe, strikis2021lb, endo2021hq, suzuki2022qe, takagi2022fl}. 
Our study focuses on the well-established ZNE technique.

ZNE was simultaneously introduced and extensively demonstrated in various applications involving systems of up to 127 qubits~\cite{chen2023sd, bruzewicz2019ti}. In ZNE, a quantum program is amplified to different noise levels through gate or pulse-level stretching, which intentionally increases the program's noise. Subsequently, the amplified results are extrapolated to estimate the noiseless values. 
Formally, the program is amplified by multiple scale factors $\lambda$. When $\lambda = 1$, the program operates at its original error rate. When $\lambda$ $> 1$, additional error, such as an identity unitary combination of the original gate, is introduced. 
The results obtained at different noise levels of $\lambda$ are collected and extrapolated to $\lambda$ $= 0$, effectively reducing noise to improve the result fidelity.

Researchers have proposed various methods for amplifying noise in quantum programs. 
For instance, the approach described in~\cite{pulse_zne} stretches gate duration to the desired levels using pulse-level control. 
Under ideal conditions, this stretching does not alter the quantum system's state, but under noisy conditions, these modifications add additional errors. Moreover, as this method requires a high degree of abstract control and calibration at the pulse level in quantum computers, it is not easily implementable across most of the existing QC systems. In contrast, unitary folding~\cite{giurgica2020dz} amplifies the noise using gate-level control, which is available in all gate-based QCs. 
This technique is based on a simple concept unitary folding: replacing the unitary operation $U\rightarrow U\left(U^{\dagger} U\right)$, where $\left(U^{\dagger} U\right)$ is an identity. 
In an ideal scenario, this replacement remains a $U$ operation, but errors corresponding to U in the quantum system amplify the errors in the $U$ operation within the modified circuit. 
Giurgica-Tiron et al. ~\cite{giurgica2020dz} proposed unitary folding with a fold from the left (which folds each unitary gate independently) and with random folding methods (which randomly select a subset of individual gates randomly as a block of unitary gates and replicates the entire block with $U\left(U^{\dagger} U\right)$. 
This method requires no knowledge of the QC underlying noise model. Although these scaling methods effectively amplify noise in the original input quantum program, they neglect the sources of error imbalance within the QC system, potentially leading to biased results and errors that may compromise the accuracy of the extrapolation results.

To address this issue, we introduce a noise-aware folding method that adjusts the quantum circuit, aiming to balance the error rate across all logical qubits by leveraging the calibration data noise model, which is freely available on most of the QC systems. 
Our approach comprehensively analyzes the calibration data noise model, strategically adjusts the scaling factor for each gate operation, balancing the distribution of gate error rates across all logical qubits. 
Addressing the error source imbalances mitigates bias-induced errors, thereby enhancing the accuracy of ZNE extrapolated results from the quantum program.

The contributions of our study are highlighted below.
\begin{itemize}
    \item We propose a noise-aware folding method, tailored to redistribute noise across quantum circuits, thereby enhancing the accuracy of ZNE by addressing the inherent error variations within quantum systems.
    \item We seamlessly integrate our approach with gate-based computation models and noise-adaptive compilation using hardware noise models, providing versatility, mitigating bias-induced errors, and enhancing the reliability of quantum computations.
    \item We conduct rigorous experiments across various full-noise models and real quantum computers, showcasing the method's consistent performance in simulations, highlighting the challenges of scaling to larger circuits, and emphasizing the disparity between simulated and real quantum computer executions.
\end{itemize}

The remainder of this paper is organized as follows. Section~\ref{sec:background} discusses the background and related QEM methods using ZNE techniques, which motivated our research. 
Section~\ref{sec:noise_aware_folding} explains the noise-aware folding method, along with its compilation and execution scheme. 
Section~\ref{sec:results} analyzes the performance results of our approach and previous methods. 
Analytical results are further discussed in Section~\ref{sec:discussion} and conclusions are presented in Section~\ref{sec:conclusion}.

\section{Background}\label{sec:background}
This section briefly provides the fundamental concept to understand our proposed method. We present QEM, noise extrapolation model, and noise scaling methods.
\subsection{\label{sec:zne} Zero-noise Extrapolation Method}

Two prominent QEM methods were introduced by Kristan Temme et al.~\cite{temme2017em} in the same letter. The first method involves extrapolation to the zero-noise limit (known as ZNE), while the second method is known as Probabilistic Error Cancellation (PEC). Both techniques aim to enhance the quality of short-depth quantum circuits and rely on the assumption that a certain noise parameter, combined with system size and circuit depth, can be treated as a small number. In terms of noise tolerance, PEC is more robust than ZNE, but it also requires more resources, which increases exponentially with circuit scale. As a result, ZNE is more suitable for short-term quantum computers, such as those in the NISQ era.

Giurgica-Tiron et al.~\cite{giurgica2020dz} outlined the fundamental structure of the ZNE problem in their work. The framework assumes that the outcome of a quantum computation is represented by a single expected value denoted as $E(\lambda)$, where $\lambda$ signifies the noise scaling factor. This expected value could emerge from either a solitary quantum circuit or a blend of quantum circuits coupled with classical post-processing. Notably, $E(\lambda)$ is a real number, ideally estimable only under infinite measurement samples. However, in practical scenarios involving a finite number of samples $N$, solely a statistical approximation of the expected value is feasible:
\begin{equation}
    \hat{E}(\lambda) = E(\lambda) + \hat{\delta}
    \label{eq:zne_estimation}
\end{equation}
Here, $\hat{\delta}$ stands for a stochastic variable with zero mean and variance $\sigma^{2}$. While $\hat{E}(\lambda)$ is the estimator for $E(\lambda)$, put it simply $\hat{E}(0)$ for $E(\lambda =0)$ is the estimate of the ZNE with its bias and variance. We can draw a genuine prediction $y$ from the probability distribution:
\begin{equation}
    P(\hat{E}(\lambda) = y) = N(E(\lambda) - y, \sigma^{2})
\end{equation}
In this distribution, $N(\mu,\sigma^{2})$ represents a typical distribution, often Gaussian, characterized by a mean $\mu$ and variance $\sigma^{2} = \sigma_0^{2}/N$. Given a series of $m$ scaling parameters $\lambda = (\lambda_1, \lambda_2,..., \lambda_m)$, where each $\lambda_j$ is greater than or equal to 1, and the corresponding outcomes:
\begin{equation}
    y = {y_1, y_2,..., y_m}
\end{equation}

Besides the foundational structure ZNE methods, the literature has showcased various models, including linear extrapolation, exponential extrapolation, Richardson extrapolation, among others~\cite{giurgica2020dz}. In our paper, we will delve into the specifics of the linear extrapolation model, highlighting its utility and presenting experimental results. 

Linear extrapolation method stands out as one of the simplest techniques, representing a specific instance within the broader scope of polynomial extrapolation. This method relies on a polynomial model of degree $d$, expressed as:
\begin{equation}
    E_{poly}^{d} = c_0 + c_1\lambda + ... c_d\lambda^d
\end{equation}
where $c_0, c_1, ..., c_d$ denote $d+1$ unknown real parameters. From this polynomial model, the linear extrapolation variant emerges:
\begin{equation}
    E_{linear}(\lambda) = E_{poly}^{(d=1)}(\lambda) = c_0 + c_1\lambda
\end{equation}
Within this context, a straightforward analytical solution exists, manifesting as the ordinary least squared estimator of the intercept parameter:
\begin{equation}
    \hat{E}_{linear}(0) = \hat{y} - \frac{S_{\lambda y}}{S_{\lambda\lambda}}\hat{x},
\end{equation}
\begin{equation}
    \hat{\lambda} = \frac{1}{m}\sum_j{\lambda_j}, \hspace{15mm} \hat{y} = \frac{1}{m}\sum_j {y_j}, 
\end{equation}
\begin{equation}
    S_{\lambda y} = \sum_j (\lambda_j - \hat{\lambda})(y_j - \hat{y}), \hspace{15mm} S_{(\lambda\lambda)} = \sum_j{(\lambda_j - \hat{\lambda})}^2
\end{equation}

where $\hat{\lambda}$ and $\hat{y}$ represent the averages of the scaling parameters $\lambda_j$ and the corresponding outcomes $y_j$, respectively. The terms $S_{\lambda y}$ and $S_{\lambda\lambda}$ are computed based on the deviations of $\lambda_j$ from their mean $\hat{\lambda}$ and the subsequent squares, facilitating the derivation of the intercept parameter. This analytical solution underscores the efficiency of the linear extrapolation method, offering a streamlined approach to predict outcomes $y$ with respect to input scaling parameters $\lambda$.

\subsection{\label{sec:noise_scaling_methods} Noise Scaling Methods}

\begin{figure}[t]
    \centering{\includegraphics[width=130mm]{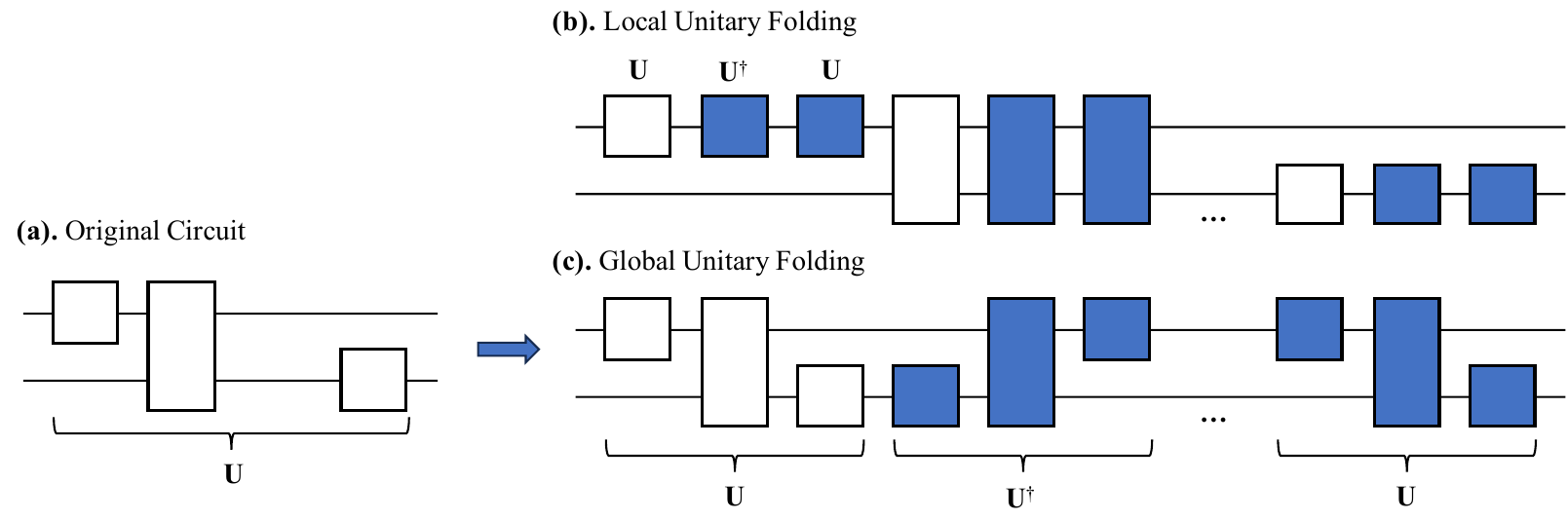}}
    \caption{Folding methods for ZNE are illustrated with the input circuit depicted in (a), followed by the process of the local unitary folding method to the original circuit in (b), and the global unitary folding method in (c). Noticing the difference in the folding process, where $U^{\dagger}U$ was applied: (b) applies each operation to each of the quantum gates individually, while (c) mirrors the entire circuit as a single $U$ operation.}
    \label{fig:identity_insertion}
\end{figure}

ZNE method measures a given observable at varying levels of noise. Using the measured dependence on the noise, they then extrapolated the result to the expected noiseless value. We describe the noise scaling method using two main types of folding namely local unitary folding and global unitary folding: 
\begin{itemize}

    \item \textbf{Unitary folding} modifies the quantum circuit by incorporating additional operations or `folds', such as global folding and local folding, which counteract the effects of noise~\cite{pascuzzi2022ce}. This strategically introduces corrective operations that enhance the fault-tolerance quantum computation, ultimately mitigating the impact of errors.
    \begin{equation}
        U \rightarrow U\left(U^{\dagger} U\right)^n, \quad n = 0, 1, 2, \ldots
        \label{eq:unitary_folding}
    \end{equation}
    In the absence of noise, the replacement rule preserves the operation, as $\left(U^{\dagger} U\right)$ equals the identity. However, if noise affects $U$, the unitary folding operation roughly scales the noise by an odd integer factor $\lambda = 1 + 2n$.

    \item \textbf{Global unitary folding:} Assume that the circuit is composed of $d$ unitary layers:
    \begin{equation}
        U = L_d,...,L_2,L_1,
        \label{eq:composed_circuit}
    \end{equation}
    where $d$ represents the depth of the circuit, with each block $L_j$ capable of representing a single layer of operations or a solitary gate. Within the global folding approach, the substitution rule from Equation~\ref{eq:unitary_folding} extends across the entire circuit, as depicted in Figure \ref{fig:identity_insertion}(c). This involves duplicating the entire circuit and adding two additional unitary operations $U$, consolidating all gates within. To refine scaling, a final folding can be confined to a subset of the circuit, typically its last $s$ layers. Consequently, the general replacement rule for global folding can be summarized as follows.
    \begin{equation}
        U \rightarrow U(U^{\dagger}U)^nL_d^{\dagger}L_{d-1}^{\dagger}...L_s^{\dagger}L_s...L_{d-1}L_d
        \label{eq:local_folding}
    \end{equation}
    The total layer count of the modified circuit is $d(2n+1)+2s$, allowing a stretch of the circuit's depth to be extended up to a resolution scale of $2/d$. This allows for the application of the scaling $d \rightarrow \lambda d$, where 
    \begin{equation}
        \lambda = 1 + \frac{2k}{d}, k = 1,2,3,....
    \end{equation}
     Conversely, given any real $\lambda$, one can determine the closest integer $k$ to $d(\lambda-1)/2$, perform integer division of $k$ by $d$ to obtain $n$ and $s$, and then apply $n$ integer foldings along with a final partial folding as detailed in Equation~\ref{eq:local_folding}.

    \item \textbf{Local unitary folding:} Rather than applying global folding to a quantum circuit by appending folds at the end, an alternative approach involves folding a subset of individual gates or layers.
    As depicted in Figure~\ref{fig:identity_insertion}(b), each gate undergoes unitary folding individually using Equation~\ref{eq:unitary_folding}. 
    This results in an evident scaling of the initial number of gates or layers, $d$, by an odd integer, $1 + 2n$. 
    Similar to global folding, a final partial folding operation can be added to achieve a more finely grained scaling factor. 
    To implement such "partial" folding, let's define an arbitrary subset, $S$, of the full set of indices ${1,2,...,d}$, where the number of elements in $S$ is a specified integer, $s = |S|$. 
    Within this framework, the following gate folding rule can be defined:

    \begin{equation}
        \forall_j \in \{1,2,...,d\}, L_j \rightarrow \left\{\begin{array}{lr}
             L_j(L_j^{\dagger}L_j)^n, & if~j \notin S  \\
             L_j(L_j^{\dagger}L_j)^{n+1}, & if~j \in S
        \end{array}\right.
    \end{equation}

    The selection of elements within the subset $S$ determines the addition of noise channels at various positions throughout the circuit, resulting in various results. In the following section, we will explore the expansion of the "fold from left" rule into a noise-aware folding mechanism. This approach involves incorporating error rate data to strategically add layers of gates from the left, deviating from the previously proposed method of folding gates a fixed number of times.

\end{itemize}
\begin{figure*}[!t]
    \centering
    \includegraphics[width=1 \textwidth]{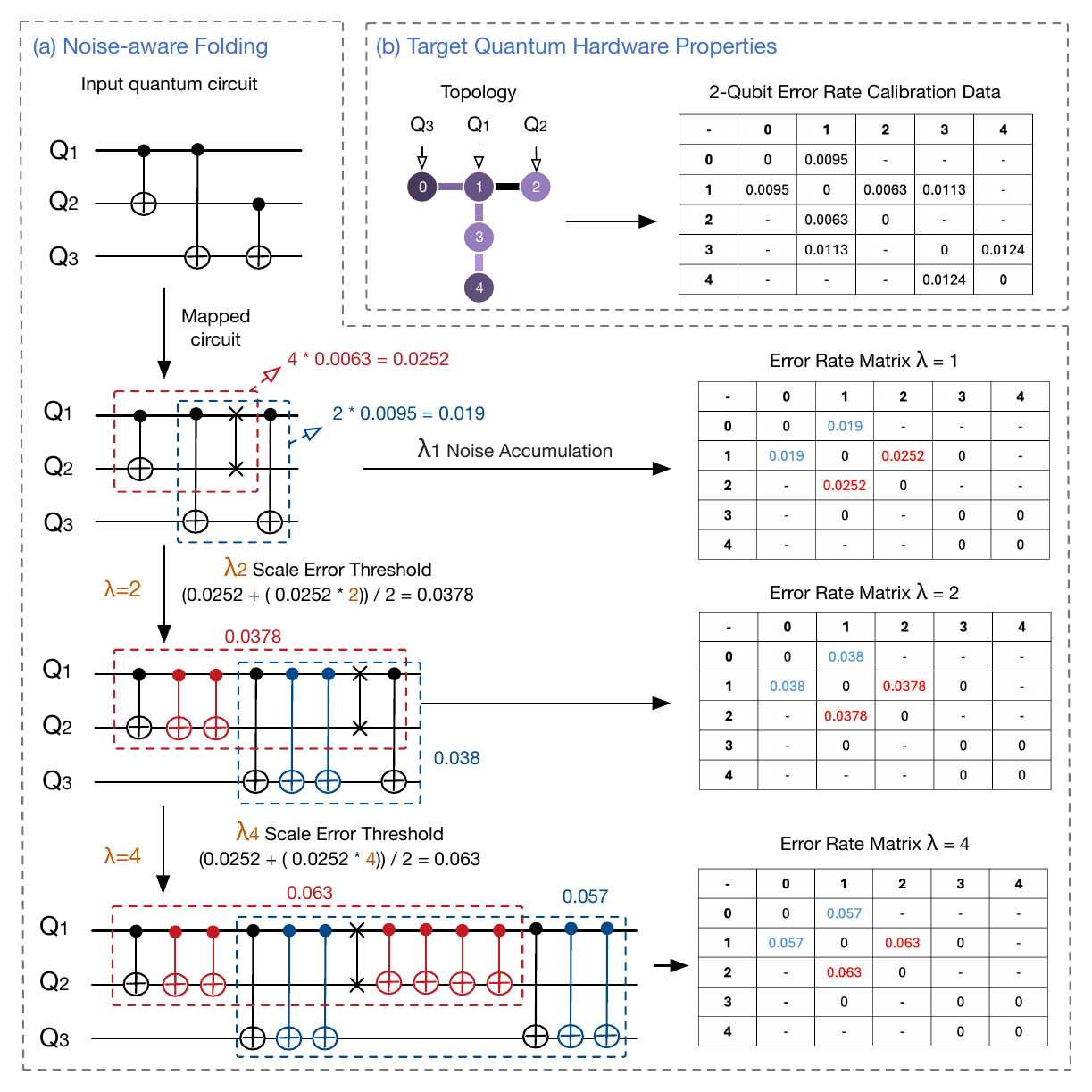}
    \caption{\label{fig:naf_overall_process} The entire process of employing noise-aware folding is described in (a), while (b) illustrates the input data, which represents the target quantum hardware. Our method begins with the application of a traditional qubit mapping process using the input quantum circuit in (a) to generate a mapped circuit for the target hardware topology in (b). Subsequently, the mapped circuit is subjected to the target hardware error rate, which facilitates the accumulation of noise. Following this, the noise-aware folding method is implemented. Initially, the noise on the circuit is accumulated to form an error rate matrix for $\lambda = 1$, representing the original mapped circuit. Then, the scaling is applied by computing the scale error rate threshold using the value $\lambda$. Finally, identity gates are inserted until the error rate of the qubit pair approaches the threshold.}
\end{figure*}
\newpage
\section{Proposed Noise-aware Folding ZNE}\label{sec:noise_aware_folding}
\subsection{Overview}\label{sec:overview}

As previously outlined, we adjusted the gate count of the quantum circuit using the scale factor ($\lambda$) through gate-level unitary folding. However, additional techniques such as identity gate insertion~\cite{he2020zn} and pulse stretching~\cite{pulse_zne} are also utilized to accommodate noise scaling.

Although some of these approaches require no specific hardware noise models, they assume uniform noise scaling across the circuit, ignoring the variations in error levels among qubits within the quantum system. In real systems, where error rates differ at different logical gates for each physical qubit, these scaling methods cannot obtain uniform scaling factors. Such nonuniform scaling factors can impede the convergence of the extrapolation model toward a zero-noise state effectively.

To address these problems, we propose to leverage the noise model of the target hardware to scale the noise using a noise-aware folding method in ZNE for the quantum circuit. 
Our approach regards $\lambda$ not solely as a gate-count scaling factor which is the main differ from fold from left method, but as an error-rate modifier threshold to scale the quantum circuit. Quantum computers require periodic calibration, which means that calibration data are readily available to the public. Although the calibration data is limited to a certain logical gates such as (one- and two-qubit gate) especially the native gate of the quantum computer, our proposed method can be used with the available data as little as a calibration error rate of a single logical two-qubit gate.
Our approach optimizes the mapping using the noise-adaptive compilation proposed by Murali et al.~\cite{noise_adaptive}, which favors qubits with high resilience. This method maps logical qubits to physical qubits with low noise levels and minimal distance, thus reducing the SWAP gate requirements.
Employing our noise-aware folding algorithm, we then effectively scale the approximate error rates within the mapped circuit. The following section explains the specifics of our noise-aware folding method.

\subsection{Noise Accumulation}\label{sec:noise_accumulation}

Figure~\ref{fig:naf_overall_process} depicts an input quantum circuit alongside the target quantum hardware properties, encompassing the qubit connectivity topology and the calibration data for 2-qubit error rates. In this illustration, we focus on the two-qubit gate due to its higher error rate compared to a one-qubit gate. Nevertheless, our method is readily adaptable to support both one- and two-qubit gates.
This visualization helps to understand how the circuit will be mapped onto the hardware and how errors may affect its execution. 
The error rate matrix contains a 2-qubit error rate for all of the qubits that have a connection to another qubit.
First, the input quantum circuit undergoes noise-adaptive compilation, as described in Section~\ref{sec:noise_aware_folding}, efficiently mapping the quantum circuit onto the target hardware by selecting the most robust qubits with low error rates, shown in Figure~\ref{fig:naf_overall_process}. 
In this process, logical qubits $Q_1$, $Q_2$, and $Q_3$ are mapped to physical qubits $1$, $0$, and $2$, respectively. 
These specific physical qubits were chosen because their error rates for 2-qubit gates are lower compared to the rest of the qubits in the system.
Although noise is intentionally added to the circuit in the next process of all of the folding methods, noise-adaptive mapping is crucial for ensuring reliable operations, minimizing qubit movement, and reducing the need for SWAP gates, which largely affect the performances of folding methods. 
To leverage this advantage, a deeper understanding of quantum computer noise—often overlooked in previous ZNE researches is crucial. Let's consider a mapped circuit comprised of $d$ unitary layers. However, unlike Equation~\ref{eq:composed_circuit} where each layer $L$ represents a unitary gate, here $L$ can be either a one- or two-qubit gate. Thus, the circuit can be represented as
\begin{equation}
    U = L_d...L_2L_1, 
\end{equation}
where for each $j \in {1,2,...,d}$, $L_j$ is defined as:
\begin{equation}
    L_j \rightarrow \left\{\begin{array}{lr}
             L_j \rightarrow g_{i,k}, & if~L_j \textit{is a two-qubit gate}  \\
             L_j \rightarrow g_{i}, & if~L_j \textit{is a one-qubit gate}
    \end{array}\right.
\end{equation}
Here, $i$ and $k$ represent logical qubits, where $g_{i,k}$ denotes a two-qubit gate acting on the control qubit $i$ and the target qubit $k$, and $g_i$ represents a one-qubit gate acting solely on the target qubit $i$.

The mapped circuit accumulates noises into an $n$ $\times$ $n$ matrix denoted as $qc\_matrix$ in Algorithm \ref{alg:noise_aware_folding}, where $n$ represents the number of qubits. In the matrix, the number of rows and columns represents the physical qubit. 
Along each off-diagonal (where $i$ and $k$ are the rows and column indices, respectively), we aggregate the noise for two-qubit gates $g_{i,k}$ between each qubit pair (where $i$ and $k$ denote the indices of the control target qubit, respectively, with $i \neq k$) by simply summing the total error rate value using the input calibration error rate to the number of two-qubit gate $g_{i,k}$. 
The noise accumulation in this phase represents $\lambda = 1$, indicating that the folding methods do not introduce any additional gates or errors to the circuit.
The error rate for each pair of qubits $i$ and $j$, the error rate is aggregated only if these qubits are physically connected and if two-qubit gates exist between them. 
While along each diagonal in the $qc\_matrix$, one-qubit gate $g_i$ are accumulated. However in our example, in Figure~\ref{fig:naf_overall_process} there are no one qubit-gate, thus making the value along the diagonal all $0$.

Once the error rates of the quantum circuit are aggregated into the error rate matrix, the circuit can be scaled at each scale factor.
Subsequent noise-aware folding processes for $\lambda$ are discussed in the next section.
\RestyleAlgo{ruled}
\SetKwComment{Comment}{/* }{ */}

\begin{algorithm} [!t]
    \KwIn{\\
        $circuit$: input transpiled circuit,
        
        $scale\_factors$: a list of scale factors $\lambda$,
        
        $backend$: target backend quantum computer\\
    }
    \KwOut{\\
        $folded\_circuits$ : a list of scaled circuits
    }
    \SetKwProg{generate}{Function \emph{}}{}{end}

    \generate{fold\_noise\_aware}{

        $cx\_error\_dict$ $\leftarrow$ GetErrorRateFrom($backend$);
        
        $qc\_matrix$ $\leftarrow$ AccumulateCircuitErrorRate($circuit$);
        
        $folded\_circuit$ $\leftarrow$ empty list; 
        
        \Comment{scale the circuit in every $\lambda$ scale factor}
        \For{$scale$ \textbf{in} $scale\_factors$}{

            $scaled\_matrix$ $\leftarrow$ $qc\_matrix$ * $scale$;
            
            $qc\_rate$ $\leftarrow$ GetHighestRateIn($qc\_matrix$);
            
            $scaled\_rate$ $\leftarrow$ GetHighestRateIn($scaled\_matrix$)

            $adjust\_rate$ $\leftarrow$ ($qc\_rate$ + $scaled\_rate)$ $/$ $2$;
            
            \Comment{Scale each pair of the qubit on the upper diagonal of the $qc\_matrix$}
            \For{$i \gets 0$ \textbf{to} $circuit.num\_qubits$}{
            
                \For{$j \gets $ $i$ \textbf{to} $ circuit.num\_qubits$}{
                    
                    $cur\_rate$ $\leftarrow$ $qc\_matrix$[i][j];
                    
                    \Comment{Perform unitary folding until the adjust rate is reached}
                    \While{$cur\_rate < adjust\_rate$}{
                        
                        \Comment{Get the qubit error rate using control (i) and target (j) qubit as index.}
                        $gate\_rate$ $\leftarrow$ $cx\_error\_dic[i,j]$;
                        
                        \Comment{For every unitary folding insertion, twice the gate $(U)$ error rate is added for $U^\dagger U$}
                        $cur\_rate$ += $gate\_rate * 2$;
                        
                        \Comment{Apply CNOT gate folding to pair i,j}
                        $circuit.cx(i, j)$;
                        
                        $circuit.cx(i, j)$;
                    }
                }
            }
            $folded\_circuit$.insert($circuit$)  
        }
    
    \Return $folded\_circuit$;
    }
    \caption{Noise-aware Folding}
    \label{alg:noise_aware_folding}
\end{algorithm}
\newpage 
\subsection{Noise-aware Folding}\label{sec:noise_scaling}

As previously discussed, $\lambda$ in our methodology is leveraged to amplify the noise accumulation and is not directly correlated with the circuit's gate count or depth as previously discussed on global or local unitary folding. 
To achieve this, we propose a noise scaling based on the error rate with the adjustment using $\lambda$ as follow:

\begin{equation}
    \epsilon_{max} = (\epsilon_{circuit} + \epsilon_{\lambda})/\gamma,
\end{equation}
where $\epsilon_{circuit}$ is a real number representing the highest error rate in the matrix accumulated at $\lambda = 1$, as shown in Figure~\ref{fig:naf_overall_process}. Here, $\epsilon_{\lambda}$ is the error rate adjusted by $\lambda$, calculated as $\epsilon_{\lambda} = \epsilon_{circuit} * \lambda$. 
Finally, $\gamma$ is an adjustable coefficient parameter that allows control the rate of the scaling. In Figure~\ref{fig:naf_overall_process} and algorithm~\ref{alg:noise_aware_folding}, $\gamma$ is set to $2$. This computation is detailed in the algorithm~\ref{alg:noise_aware_folding} line 6-9.

Using $\epsilon_{max}$ at any $\lambda$, our noise-aware folding mechanism applies the unitary folding using Equation~\ref{eq:unitary_folding} until the folded gate error accumulation reaches close or equal to $\epsilon_{max}$, as detailed in algorithm~\ref{alg:noise_aware_folding} line 13-17.

This approach is noise-aware and significantly differs from previous methods because the rate at which unitary folding is applied to the circuit depends entirely on the error rate of the physical qubit. 
As illustrated in Figure~\ref{fig:naf_overall_process}, at $\lambda = 2$, unitary folding was applied equally by adding 2 CNOT gates to pair of logical qubit $Q_{1,2}$ and $Q_{2,3}$ on the quantum circuit. 
However notice at $\lambda = 4$, 4 CNOT gates were added to the pair of logical qubit $Q_{1,2}$ while only 2 CNOT gates were added to the pair of logical qubit $Q_{2,3}$ in the quantum circuit despite pair of $Q_{1,2}$ has more CNOT gates and a SWAP gate as well. 
This is because the accumulated error rate on the pair of qubit $Q_{1,2}$ and $Q_{2,3}$ are 0.063 and 0.057, respectively, with the $\epsilon_{max} = 0.063$. $Q_{1,2}$ reaches the maximum threshold, while $Q_{2,3}$ reaches close to maximum and cannot add additional $U^{\dagger}U$ anymore due to it would be exceed $\epsilon_{max}$ and such condition is not allowed.

Note that this study focuses solely on applying the folding method for two-qubit gates due to their notable error rates and the availability of calibration data. However, our proposed noise-aware folding method can be expanded to accommodate all gate types, assuming sufficient error rate calibration data are available within the noise model.
\section{Evaluation}\label{sec:results}
This section compares the results of our noise-aware folding method with those of other extrapolation models, including the existing method referenced in \cite{giurgica2020dz}, using the linear fit extrapolation model.
\begin{figure}[t]
    \centering{\includegraphics[width=55mm]{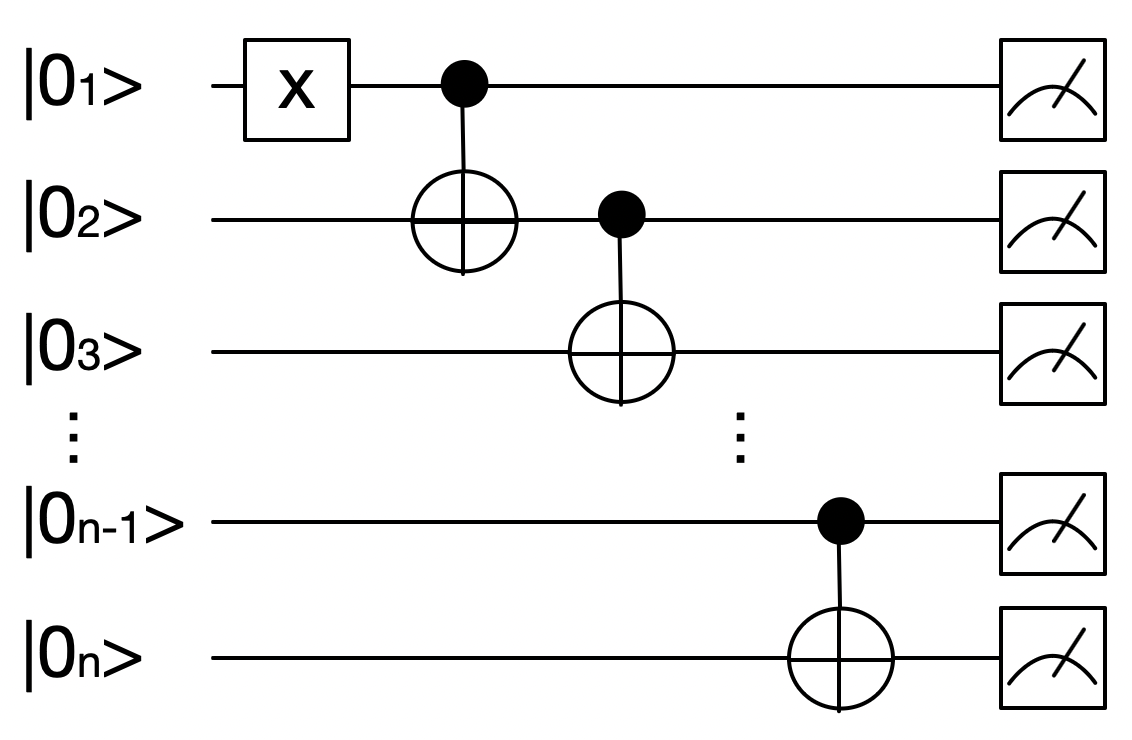}}
    \caption{The circuit configuration designed to benchmark ZNE using various folding methods begins with initializing each qubit to the state $|0\rangle$. Subsequently, a NOT (X) gate is applied to the first qubit, followed by successive CNOT gates between adjacent qubits (Qubit 1 and 2, 2 and 3, ..., up to n-1 and n). This sequence results in all qubits being measured in the state $|1\rangle$.}
    \label{fig:benchmark_circuit_cnot}
\end{figure}
\subsection{Experimental Setup}\label{sec:setup}
\begin{figure*} 
    \captionsetup[subfigure]{justification=centering}
    \begin{subfigure}[t]{1\columnwidth}
        \centering
        \includegraphics[width=\columnwidth]{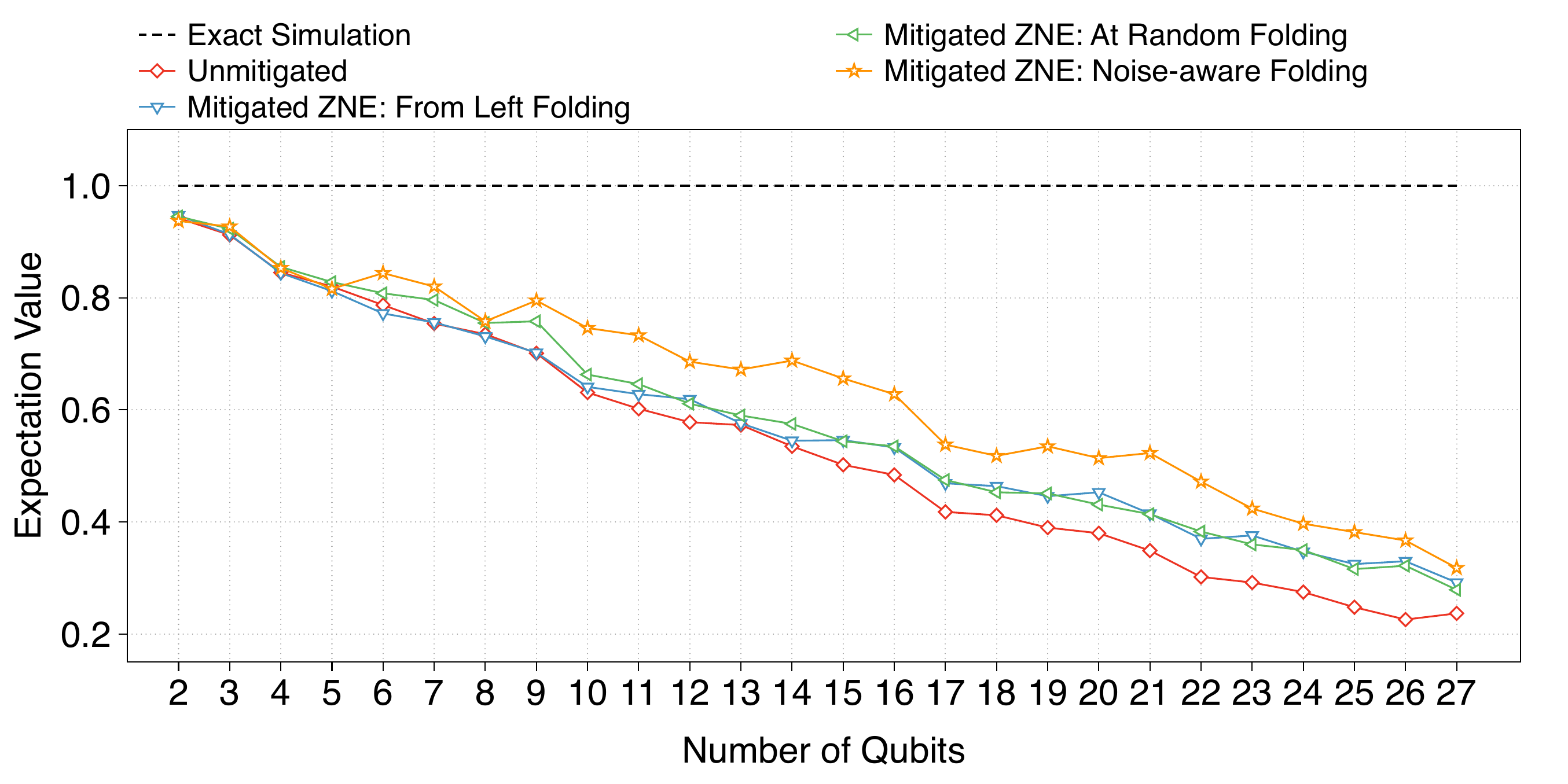}
        \caption{\textit{ibm\_mumbai} noise model result.}
        \label{fig:eval_fake_mumbai}
    \end{subfigure}
    \hspace{\fill}
    \hfill
    \begin{subfigure}[t]{1\columnwidth}
        \centering
        \includegraphics[width=\columnwidth]{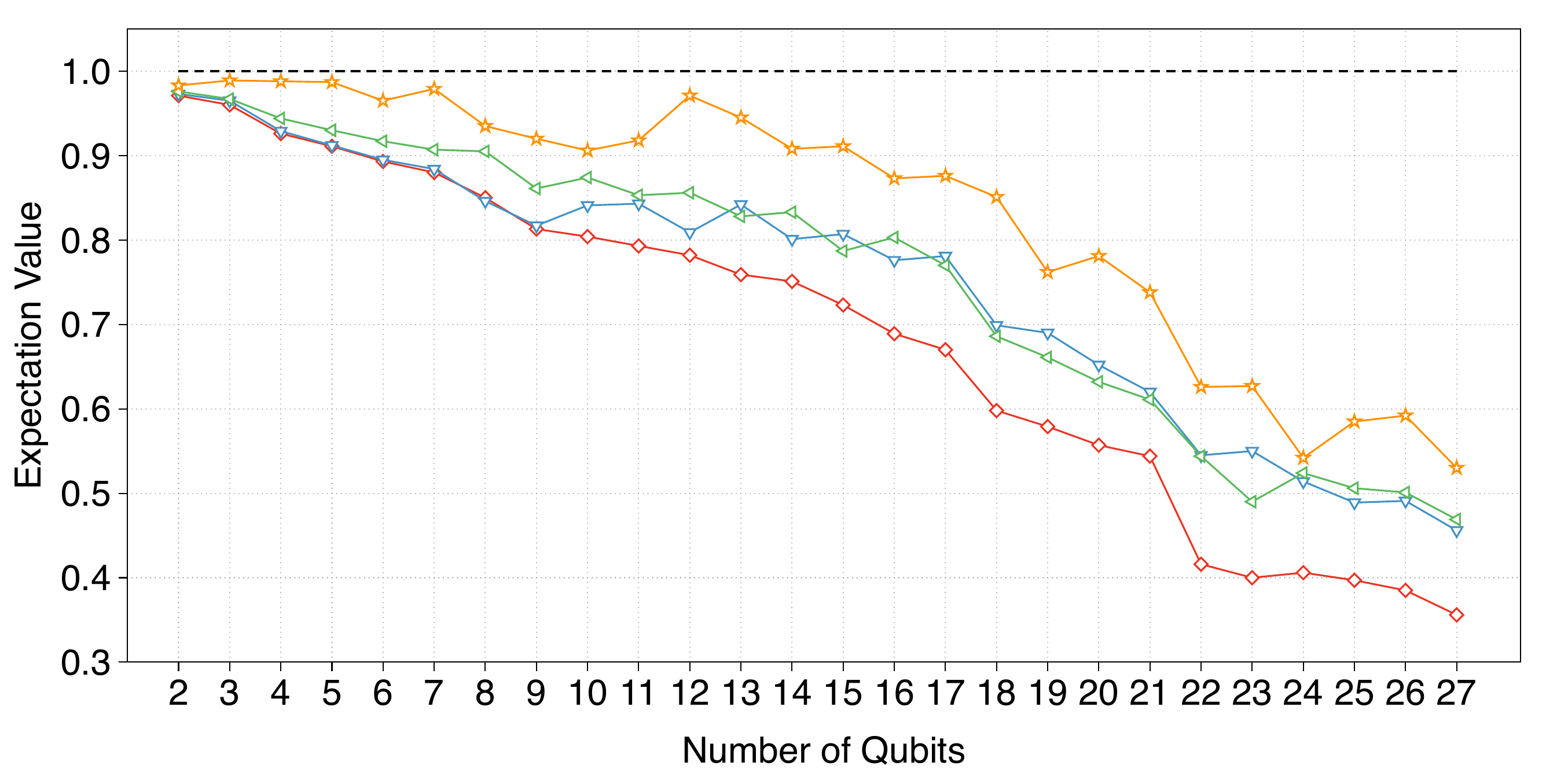}
        \caption{\textit{ibm\_cairo} noise model result.}
        \label{fig:eval_fake_cairo}
    \end{subfigure}
    \caption{Comparison of expectation values (vertical axis) obtained through using ZNE linear fit model with different folding methods on a quantum computer simulator, each with its respective noise model. (a) and (b) show results using noise models from \textit{ibm\_mumbai} and \textit{ibm\_cairo}, respectively. The evaluation involves three folding methods: fold from left, fold at random, and noise-aware folding applied to the circuit depicted in Figure~\ref{fig:benchmark_circuit_cnot}, spanning from 2 to 27 qubits (horizontal axis). The scaling factor $\lambda$ of $[1, 1.5, 2, 2.5]$ was utilized for each folding method at every qubit count. The `Exact Simulation' represents results without any applied noise model, while the `unmitigated' results showcase the circuit execution using a noise model without error mitigation. Each result is represented by a distinct symbol and color, as shown in (a).}
    \label{fig:cnot_benchmark_result}
\end{figure*}
Our proposed method was validated in various full-noise models from quantum computers available on Qiskit~\cite{Qiskit}. 
In these evaluations, we evaluated both the setup circuit in Figure~\ref{fig:benchmark_circuit_cnot} and the Bernstein-Vazirani (BV) circuit as a benchmark~\cite{bernstein}. 
We selected this circuit (see Figure \ref{fig:benchmark_circuit_cnot}) to assess our proposed method due to its similarity to the fully-entangled layer utilized in various NISQ applications.
At a logical level, it presents a relatively straightforward circuit that alters the states of individual qubits. 
When errors or noise are introduced to this circuit, they are non-negligibly propagated between qubits and become significant due to their correlation and the necessity to change subsequent qubit states. 
Any error during execution notably impacts the subsequent qubit within the circuit, depending on the point of its incidence. 
Moreover, owing to the connectivity constraints, this circuit is not easily executable on quantum computers. 
Accommodating these limitations in quantum computer connectivity becomes increasingly complex with the increasing number of qubits.

For each noise model, we ran the quantum circuit with different scaling parameters for comparison purposes. Along with our noise-aware folding method, we employed both from-left and at-random folding methods in \cite{giurgica2020dz} implemented on the Mitiq framework \cite{LaRose_2022}.
However, owing to the nature of ZNE, we must preserve the folded gates in the folded circuits, which precludes optimization such as gate cancelations before executing the quantum circuit at the compilation level. 
Given the complexity of our benchmark circuit (Figure~\ref{fig:benchmark_circuit_cnot}) and the connectivity limitations of the quantum computer, we must introduce numerous SWAP gates as the number of qubits increases. 
To avoid this problem and ensure a fair comparison, we preemptively transpile the quantum circuit using Qiskit at the highest $optimization\_level = 3$ before initiating the folding process in each method. 
The resulting transpiled circuit was input to each folding method, each with its own processes for extrapolation.

In addition, all experimental circuits were executed on real quantum computers (see Subsection~\ref{sec:real_qc_result}).
All comparison methods, both on simulators and on real quantum computers, employed the same parameters and compilation process. 
The subsequent subsection delves into the evaluation results.
\begin{figure}
\centering{\includegraphics[width=0.85\columnwidth]{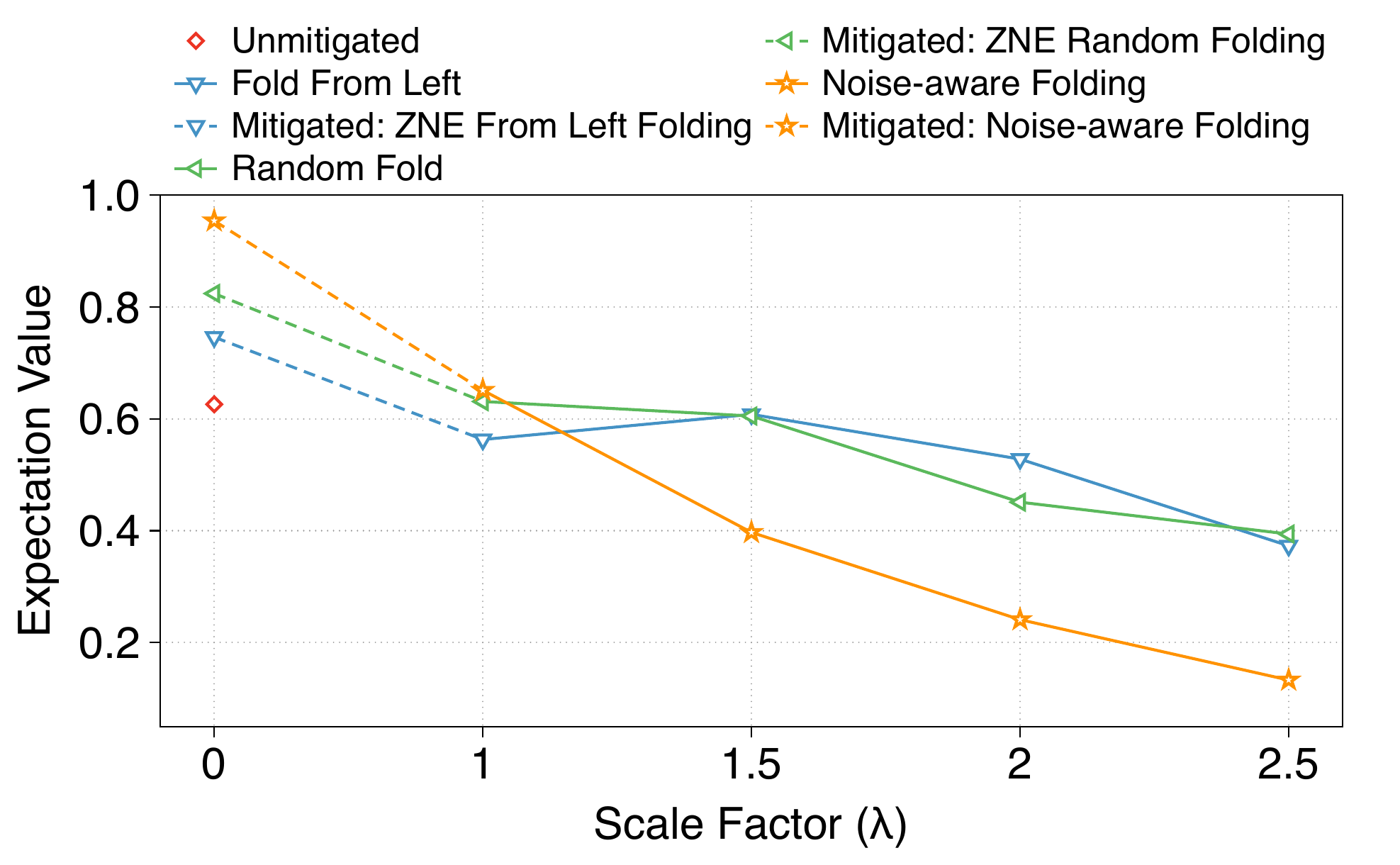}}
    \caption{Comparison of the expectation value using different folding methods on a Bernstein-Vazirani (BV) \cite{bernstein} circuit with 14-qubit on \textit{ibm\_cairo} noise model simulation. The scale factor shows the drop in expectation value in different folding methods as the circuit gets nosier, while $\lambda = 0$ is the extrapolated expectation value where zero noise is assumed.}
    \label{fig:bv_n14_benchmark}
\end{figure}
\subsection{Result on Simulators}\label{sec:sim_qc_result}

Our experiments were conducted on the 27-qubit system, making use of the full-noise models implemented in Qiskit.
Each experiment was conducted five times with different qubit counts and folding methods. 
The results are the average of the five experiments for each folding method. The following graphs present the result of the linear fit extrapolation model, which demonstrated notably reliable results across all qubit counts. 

Figure~\ref{fig:cnot_benchmark_result} depicts the results of executing the benchmark circuit shown in Figure~\ref{fig:benchmark_circuit_cnot} on two full-noise models, namely, \textit{ibm\_mumbai} and \textit{ibm\_cairo}, using a simulator. 
As shown in panels (a) and (b) of this figure, the expectation decreased with increasing qubit counts. Applying the unmitigated, fold from left, random folding, and noise-aware folding to the benchmark circuit in the \textit{ibm\_mumbai} noise model, the expectation values respectively decreased from 0.943, 0.952, 0.949, and 0.959 in the 2-qubit circuits to 0.237, 0.292, 0.279, and 0.318 in the 27-qubit circuit (Figure~\ref{fig:eval_fake_mumbai}).

The expectation values of all execution methods, (including the unmitigated method) were high in the \textit{ibm\_cairo} noise model than in the \textit{ibm\_mumbai} (c.f. Figure~\ref{fig:eval_fake_mumbai} and ~\ref{fig:eval_fake_cairo}). This discrepancy arises from differences in the noise model parameters. Specifically, \textit{ibm\_mumbai} yields a significantly higher error rate than the other models.

Figure~\ref{fig:bv_n14_benchmark} presents the experimental results of a 14-qubit BV circuit using the full-noise model simulation from \textit{ibm\_cairo}. The expectation value reduced with increasing scaling factor $\lambda$ in our proposed method, but did not show a consistent drop in the left and random folding methods. This result demonstrates that in our proposed method, the extrapolation model accurately extrapolates to the zero-noise value.
\begin{figure}
    \centering{\includegraphics[width=1\columnwidth]{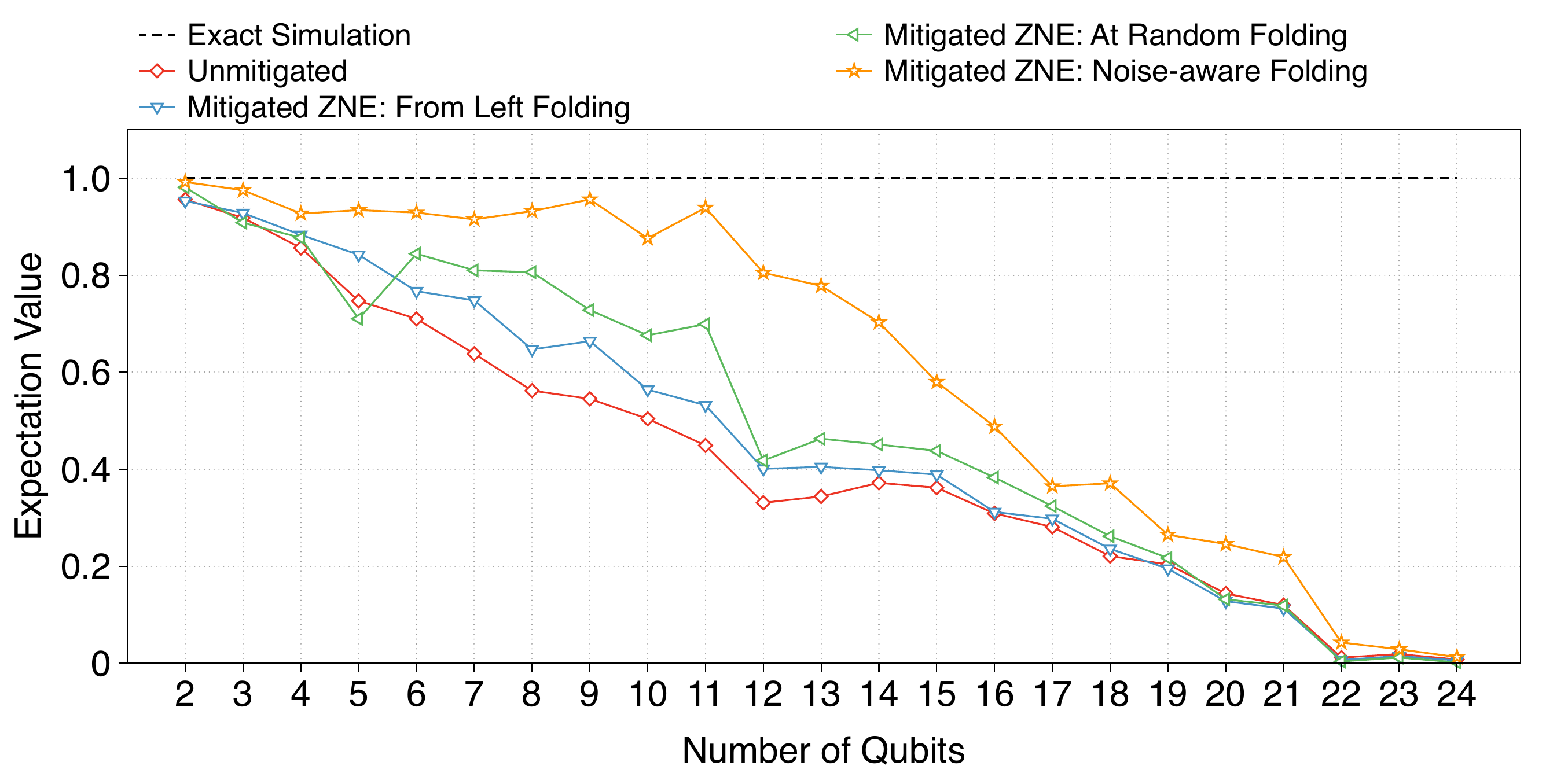}}
    \caption{Expectation value results on a real quantum computer \textit{ibmq\_mumbai} with the exact same setup and scaling factor configuration from the result in Figure~\ref{fig:cnot_benchmark_result}.}
    \label{fig:mumbai_benchmark}
\end{figure}
\subsection{Result on Real Quantum Computer}\label{sec:real_qc_result}
Figure~\ref{fig:mumbai_benchmark} presents the expectation values on the benchmark circuit shown in Figure~\ref{fig:benchmark_circuit_cnot}, obtained through experiments on a real quantum computer \textit{ibmq\_mumbai}. As of preparing this manuscript, IBM has retired \textit{ibmq\_mumbai}; however, we captured its noise model at exact same time that we performed the experiment. You can find details regarding its properties and noise model in the appendix.
The results, which mirror those of Figure~\ref{fig:cnot_benchmark_result}, were averaged over five executions of each folding method. 
Notably, the expectation values of the unmitigated, fold from left, fold at random, and noise-aware folding methods declined from 0.949, 0.956, 0.953, and 0.981, respectively, in the 2-qubit circuit to 0.007, 0.019, 0.015, and 0.12, respectively in the 24-qubit circuit. 
The proposed method consistently outperformed the existing methods from the 2 to 21 qubits.
This discrepancy can be explained by the noticeably faster decline of the expectation value on scaled circuits in our proposed method than in the existing ones. At higher $\lambda$ values with those of higher than 22-qubit, certain folded circuits reach a 0 expectation value, rendering the extrapolated results unreliable.

Comparing executions on real quantum computers to simulations, the overall results drop significantly with increasing qubit counts on real quantum computers. 
This disparity arises because the noise model cannot fully represent all noise sources in real quantum computers during simulations making the results on real quantum computers less reliable.

When the BV circuit was executed on a real quantum computer, the success rate was consistently below 1\%. 
This low success rate translates to an expectation value of less than 0.01 for the unmitigated circuit. 
Consequently, in folded circuits where the noise is increased, the success rate of the BV circuit approaches 0\% and the results become meaningless on this circuit.
\section{Discussion}\label{sec:discussion}
We applied other extrapolation methods, namely, exponential fit~\cite{endo2018pq}, adaptive-exponential fit from~\cite{giurgica2020dz}, and Richardson’s method~\cite{temme2017em}, to ZNE execution on the above-mentioned circuit folding methods. 
However, the results demonstrated no consistent substantial improvements or more reliable outcomes over those of linear fitting model. 
In fact, the models were often degraded by extreme over-fitting or under-fitting, leading to non-convergence in some instances. 
Although improvements from those of linear fit extrapolation were sometimes observed, the results of these fittings were too inconsistent for practical execution, particularly in scenarios involving variational-based quantum algorithms where a consistent result is required for many iterations. 
Notably, the exponential fit and adaptive-exponential fit methods produced identical outputs across all folding methods, including our proposed noise-aware folding approach. 
Hence, the effectiveness of these extrapolations in discerning the performances of various extrapolation techniques are uncertain.
Although linear fit extrapolation, like any other models, is unable to extrapolate to zero noise as theoretically proposed due to strong noise in the quantum computer, the consistent improvement in results we observe using linear fit sets it apart from other extrapolation models. 
However, utilizing zero-noise extrapolation comes with a trade-off of executing more quantum circuits, typically ranging from $2x$ to $5x$ the original amount, depending on the number of $\lambda$ values desired by the user.

Moreover, while QEM techniques like ZNE are primarily employed to enhance the results of NISQ devices, they may also prove practical for use in quantum error correction codes to protect qubits from errors. 
By employing both error correction codes and mitigation strategies, a more resource-efficient method can be achieved compared to traditional error correction codes alone. 
In this hybrid approach, QEM techniques can be used to pre-process the noisy qubits before encoding them into a QEC code. 
This can reduce the error rates of the qubits, making the subsequent QEC code more effective and efficient. For instance, ZNE can be applied to suppress errors in the qubits before encoding them into a surface code or a concatenated code. 
Alternatively, QEM techniques can be integrated directly into the QEC code, allowing for real-time error mitigation during the execution of the quantum algorithm. 
This can be particularly useful for codes that are prone to error accumulation, such as those used in quantum simulations or machine learning applications. 
As advancements in quantum computer technology reduce noise levels, making error mitigation strategies more effective, this potential becomes even more viable. 
With lower noise levels, QEM techniques can provide a greater reduction in error rates, further enhancing the performance of the hybrid approach. 

\section{Conclusion}\label{sec:conclusion}

We proposed a noise-aware folding method to improve the zero-noise extrapolation method by employing the noise-adaptive compilation and a folding mechanism that distributes noise uniformly using calibration data based on the quantum hardware noise model.
By dynamically adjusting the error rates based on hardware noise models, it improves the precision and reliability of the linear fit extrapolation method results in improving the quantum computer result fidelity 
The fidelity improvements over the compared methods reached 35\% and 31\% on quantum computer simulators and real quantum computers, respectively. 
Although demonstrating robustness in simulations and smaller-scale executions on real quantum computers, the proposed method was challenged by scaling to larger circuits, necessitating further exploration and optimization to broaden its applicability. 
Further refinements and adaptations are essential to enhance scalability and reliability, driving quantum error mitigation methods towards realizing error-free quantum computations.

\bmhead{Acknowledgments}
This research was partly supported by Quantum Computing based on Quantum Advantage challenge research (RS-2023-00257994) through the National Research Foundation of Korea(NRF) funded by the Korean government (MSIT) and Institute for Information \& communications Technology Planning \& Evaluation (IITP) grant funded by the Korea government (MSIT) (No. 2020-0-00014, A Technology Development of Quantum OS for Fault-tolerant Logical Qubit Computing Environment).

\newpage
\begin{appendices}
\section{ibmq\_mumbai's Properties}\label{asec:experiment setup}
This supplementary document provides the experimental setup detail of the quantum computer that we use in our demonstration.

\begin{figure}[htbp]
    \centering{\includegraphics[width=0.7\columnwidth]{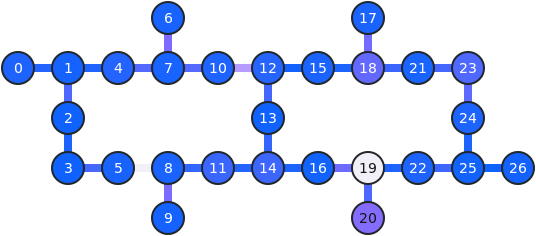}}
    \caption{\textit{ibmq\_mumbai}'s qubit connectivity topology.}
    \label{fig:mumbai_benchmark_topology}
\end{figure}

\begin{table}[!ht]
    \centering
    \begin{tabular}{|c|c|c|c|c|}
    \hline
        Qubit & T1 (us) & T2 (us) & Frequency (GHz) & Anharmonicity (GHz) \\ \hline
        0 & 111.6961464 & 179.1588076 & 5.070549646 & -0.328451763 \\ \hline
        1 & 122.2484892 & 197.5837264 & 4.929863389 & -0.331291977 \\ \hline
        2 & 81.90697735 & 155.8248129 & 4.670281392 & -0.336876757 \\ \hline
        3 & 167.7648109 & 122.141877 & 4.889402396 & -0.331179264 \\ \hline
        4 & 42.05229013 & 29.68056521 & 5.020938057 & -0.330174519 \\ \hline
        5 & 101.1535787 & 122.3569753 & 4.969532374 & -0.329757062 \\ \hline
        6 & 103.291911 & 45.33644272 & 4.965898301 & -0.329305699 \\ \hline
        7 & 67.91110372 & 48.58658623 & 4.894369882 & -0.331001624 \\ \hline
        8 & 87.45290954 & 142.5387495 & 4.79164979 & -0.332559138 \\ \hline
        9 & 96.66829381 & 73.24259076 & 4.954976802 & -0.330638393 \\ \hline
        10 & 110.9073585 & 132.0053088 & 4.958620128 & -0.330907906 \\ \hline
        11 & 88.46621356 & 202.3471754 & 4.666070269 & -0.332629168 \\ \hline
        12 & 141.6272911 & 204.5251318 & 4.74291957 & -0.332978886 \\ \hline
        13 & 117.1107884 & 151.0626282 & 4.88858331 & -0.328060893 \\ \hline
        14 & 84.32426156 & 153.5437913 & 4.780232309 & -0.332554813 \\ \hline
        15 & 97.45073523 & 90.18823679 & 4.8581701 & -0.333242963 \\ \hline
        16 & 104.390153 & 262.8877462 & 4.98035969 & -0.329920963 \\ \hline
        17 & 104.1706916 & 85.0072165 & 5.003339003 & -0.329925892 \\ \hline
        18 & 122.1904549 & 212.2866827 & 4.781291798 & -0.333097806 \\ \hline
        19 & 117.3925825 & 251.3554841 & 4.809727654 & -0.33209643 \\ \hline
        20 & 130.2027001 & 214.4300921 & 5.047734483 & -0.328048366 \\ \hline
        21 & 109.9161771 & 200.7424141 & 4.942677405 & -0.331299817 \\ \hline
        22 & 130.3648863 & 211.195153 & 4.911344811 & -0.331838149 \\ \hline
        23 & 90.39455816 & 163.0951773 & 4.892629733 & -0.331514799 \\ \hline
        24 & 141.7988532 & 29.62637317 & 4.671041833 & -0.335931163 \\ \hline
        25 & 148.610822 & 128.6915538 & 4.758538084 & -0.333550575 \\ \hline
        26 & 72.49759913 & 135.6307972 & 4.954268232 & -0.329535731 \\ \hline
    \end{tabular}
    \caption{\textit{ibmq\_mumber}'s noise model captured on 26-March-2024 from IBM.}
\end{table}

\newgeometry{margin=2cm}
\begin{landscape}
\begin{table}[!ht]
    \centering
    \begin{tabular}{|c|c|c|c|c|c|c|c|c|}
    \hline
        Qubit & Readout assignment error  & Prob meas0 prep1  & Prob meas1 prep0  & Readout length (ns) & ID error  & Sx error  & Pauli-X error  & CNOT error   \\ \hline
        0 & 2.26E-02 & 3.54E-02 & 9.80E-03 & 3512.888889 & 2.74E-04 & 2.74E-04 & 2.74E-04 & 0\_1:5.62E-03  \\ \hline
        1 & 1.58E-02 & 2.62E-02 & 5.40E-03 & 3512.888889 & 1.93E-04 & 1.93E-04 & 1.93E-04 & 1\_2:8.25E-03;1\_4:5.29E-03;1\_0:5.62E-03  \\ \hline
        2 & 1.15E-02 & 1.60E-02 & 7.00E-03 & 3512.888889 & 1.16E-04 & 1.16E-04 & 1.16E-04 & 2\_3:5.11E-03;2\_1:8.25E-03  \\ \hline
        3 & 1.15E-02 & 1.74E-02 & 5.60E-03 & 3512.888889 & 2.33E-04 & 2.33E-04 & 2.33E-04 & 3\_2:5.11E-03;3\_5:7.57E-03  \\ \hline
        4 & 2.73E-02 & 3.72E-02 & 1.74E-02 & 3512.888889 & 2.77E-04 & 2.77E-04 & 2.77E-04 & 4\_7:8.49E-03;4\_1:5.29E-03  \\ \hline
        5 & 1.74E-02 & 2.32E-02 & 1.16E-02 & 3512.888889 & 2.33E-04 & 2.33E-04 & 2.33E-04 & 5\_3:7.57E-03;5\_8:2.25E-02  \\ \hline
        6 & 1.72E-02 & 2.58E-02 & 8.60E-03 & 3512.888889 & 2.57E-04 & 2.57E-04 & 2.57E-04 & 6\_7:1.22E-02  \\ \hline
        7 & 1.88E-02 & 3.12E-02 & 6.40E-03 & 3512.888889 & 4.94E-04 & 4.94E-04 & 4.94E-04 & 7\_10:8.76E-03;7\_6:1.22E-02;7\_4:8.49E-03  \\ \hline
        8 & 1.97E-02 & 2.96E-02 & 9.80E-03 & 3512.888889 & 1.90E-04 & 1.90E-04 & 1.90E-04 & 8\_11:6.50E-03;8\_5:2.25E-02;8\_9:1.21E-02  \\ \hline
        9 & 1.52E-02 & 2.34E-02 & 7.00E-03 & 3512.888889 & 2.63E-04 & 2.63E-04 & 2.63E-04 & 9\_8:1.21E-02  \\ \hline
        10 & 2.28E-02 & 3.38E-02 & 1.18E-02 & 3512.888889 & 4.45E-04 & 4.45E-04 & 4.45E-04 & 10\_12:1.59E-02;10\_7:8.76E-03  \\ \hline
        11 & 4.44E-02 & 5.60E-02 & 3.28E-02 & 3512.888889 & 2.47E-04 & 2.47E-04 & 2.47E-04 & 11\_14:5.62E-03;11\_8:6.50E-03  \\ \hline
        12 & 2.83E-02 & 1.90E-02 & 3.76E-02 & 3512.888889 & 1.62E-04 & 1.62E-04 & 1.62E-04 & 12\_13:7.55E-03;12\_10:1.59E-02;12\_15:4.46E-03  \\ \hline
        13 & 1.16E-02 & 1.46E-02 & 8.60E-03 & 3512.888889 & 1.70E-04 & 1.70E-04 & 1.70E-04 & 13\_12:7.55E-03;13\_14:6.55E-03  \\ \hline
        14 & 4.60E-02 & 3.74E-02 & 5.46E-02 & 3512.888889 & 1.61E-04 & 1.61E-04 & 1.61E-04 & 14\_11:5.62E-03;14\_13:6.55E-03;14\_16:5.07E-03  \\ \hline
        15 & 2.01E-02 & 2.54E-02 & 1.48E-02 & 3512.888889 & 1.93E-04 & 1.93E-04 & 1.93E-04 & 15\_18:5.09E-03;15\_12:0.0044594326370586135  \\ \hline
        16 & 1.28E-02 & 2.16E-02 & 4.00E-03 & 3512.888889 & 1.94E-04 & 1.94E-04 & 1.94E-04 & 16\_19:0.01074071841628857;16\_14:5.07E-03  \\ \hline
        17 & 1.98E-02 & 3.18E-02 & 7.80E-03 & 3512.888889 & 3.48E-04 & 3.48E-04 & 3.48E-04 & 17\_18:1.14E-02  \\ \hline
        18 & 8.00E-02 & 9.62E-02 & 6.38E-02 & 3512.888889 & 1.93E-04 & 1.93E-04 & 1.93E-04 & 18\_21:6.97E-03;18\_15:5.09E-03;18\_17:1.14E-02  \\ \hline
        19 & 2.31E-01 & 2.33E-01 & 2.29E-01 & 3512.888889 & 2.49E-04 & 2.49E-04 & 2.49E-04 & 19\_22:5.13E-03;19\_20:8.25E-03;19\_16:1.07E-02  \\ \hline
        20 & 1.13E-01 & 3.26E-02 & 1.94E-01 & 3512.888889 & 1.81E-04 & 1.81E-04 & 1.81E-04 & 20\_19:8.25E-03  \\ \hline
        21 & 1.91E-02 & 2.80E-02 & 1.02E-02 & 3512.888889 & 3.70E-04 & 3.70E-04 & 3.70E-04 & 21\_18:6.97E-03;21\_23:7.72E-03  \\ \hline
        22 & 2.33E-02 & 3.96E-02 & 7.00E-03 & 3512.888889 & 1.87E-04 & 1.87E-04 & 1.87E-04 & 22\_19:5.13E-03;22\_25:7.16E-03  \\ \hline
        23 & 6.02E-02 & 7.44E-02 & 4.60E-02 & 3512.888889 & 2.62E-04 & 2.62E-04 & 2.62E-04 & 23\_21:7.72E-03;23\_24:9.50E-03  \\ \hline
        24 & 2.17E-02 & 3.36E-02 & 9.80E-03 & 3512.888889 & 3.61E-04 & 3.61E-04 & 3.61E-04 & 24\_23:9.50E-03;24\_25:5.22E-03  \\ \hline
        25 & 1.31E-02 & 1.92E-02 & 7.00E-03 & 3512.888889 & 1.42E-04 & 1.42E-04 & 1.42E-04 & 25\_22:7.16E-03;25\_26:4.25E-03;25\_24:5.22E-03  \\ \hline
        26 & 1.41E-02 & 1.90E-02 & 9.20E-03 & 3512.888889 & 1.65E-04 & 1.65E-04 & 1.65E-04 & 26\_25:4.25E-03 \\ \hline
    \end{tabular}
    \caption{Extended \textit{ibmq\_mumber}'s noise model including full detail of the gate error rate captured on 26-March-2024 from IBM.}

\end{table}
\end{landscape}
\restoregeometry





\end{appendices}


\bibliography{manuscript}


\begin{thebibliography}{36}
\ifx \bisbn   \undefined \def \bisbn  #1{ISBN #1}\fi
\ifx \binits  \undefined \def \binits#1{#1}\fi
\ifx \bauthor  \undefined \def \bauthor#1{#1}\fi
\ifx \batitle  \undefined \def \batitle#1{#1}\fi
\ifx \bjtitle  \undefined \def \bjtitle#1{#1}\fi
\ifx \bvolume  \undefined \def \bvolume#1{\textbf{#1}}\fi
\ifx \byear  \undefined \def \byear#1{#1}\fi
\ifx \bissue  \undefined \def \bissue#1{#1}\fi
\ifx \bfpage  \undefined \def \bfpage#1{#1}\fi
\ifx \blpage  \undefined \def \blpage #1{#1}\fi
\ifx \burl  \undefined \def \burl#1{\textsf{#1}}\fi
\ifx \doiurl  \undefined \def \doiurl#1{\url{https://doi.org/#1}}\fi
\ifx \betal  \undefined \def \betal{\textit{et al.}}\fi
\ifx \binstitute  \undefined \def \binstitute#1{#1}\fi
\ifx \binstitutionaled  \undefined \def \binstitutionaled#1{#1}\fi
\ifx \bctitle  \undefined \def \bctitle#1{#1}\fi
\ifx \beditor  \undefined \def \beditor#1{#1}\fi
\ifx \bpublisher  \undefined \def \bpublisher#1{#1}\fi
\ifx \bbtitle  \undefined \def \bbtitle#1{#1}\fi
\ifx \bedition  \undefined \def \bedition#1{#1}\fi
\ifx \bseriesno  \undefined \def \bseriesno#1{#1}\fi
\ifx \blocation  \undefined \def \blocation#1{#1}\fi
\ifx \bsertitle  \undefined \def \bsertitle#1{#1}\fi
\ifx \bsnm \undefined \def \bsnm#1{#1}\fi
\ifx \bsuffix \undefined \def \bsuffix#1{#1}\fi
\ifx \bparticle \undefined \def \bparticle#1{#1}\fi
\ifx \barticle \undefined \def \barticle#1{#1}\fi
\bibcommenthead
\ifx \bconfdate \undefined \def \bconfdate #1{#1}\fi
\ifx \botherref \undefined \def \botherref #1{#1}\fi
\ifx \url \undefined \def \url#1{\textsf{#1}}\fi
\ifx \bchapter \undefined \def \bchapter#1{#1}\fi
\ifx \bbook \undefined \def \bbook#1{#1}\fi
\ifx \bcomment \undefined \def \bcomment#1{#1}\fi
\ifx \oauthor \undefined \def \oauthor#1{#1}\fi
\ifx \citeauthoryear \undefined \def \citeauthoryear#1{#1}\fi
\ifx \endbibitem  \undefined \def \endbibitem {}\fi
\ifx \bconflocation  \undefined \def \bconflocation#1{#1}\fi
\ifx \arxivurl  \undefined \def \arxivurl#1{\textsf{#1}}\fi
\csname PreBibitemsHook\endcsname

\bibitem[\protect\citeauthoryear{Lau et~al.}{2022}]{lau2022nc}
\begin{barticle}
\bauthor{\bsnm{Lau}, \binits{J.W.Z.}},
\bauthor{\bsnm{Lim}, \binits{K.H.}},
\bauthor{\bsnm{Shrotriya}, \binits{H.}},
\bauthor{\bsnm{Kwek}, \binits{L.C.}}:
\batitle{Nisq computing: where are we and where do we go?}
\bjtitle{AAPPS Bulletin}
\bvolume{32}(\bissue{1}),
\bfpage{27}
(\byear{2022})
\end{barticle}
\endbibitem

\bibitem[\protect\citeauthoryear{Huang et~al.}{2020}]{huang2020sq}
\begin{barticle}
\bauthor{\bsnm{Huang}, \binits{H.-L.}},
\bauthor{\bsnm{Wu}, \binits{D.}},
\bauthor{\bsnm{Fan}, \binits{D.}},
\bauthor{\bsnm{Zhu}, \binits{X.}}:
\batitle{Superconducting quantum computing: a review}.
\bjtitle{Science China Information Sciences}
\bvolume{63},
\bfpage{1}--\blpage{32}
(\byear{2020})
\end{barticle}
\endbibitem

\bibitem[\protect\citeauthoryear{Ezratty}{2023}]{ezratty2023po}
\begin{barticle}
\bauthor{\bsnm{Ezratty}, \binits{O.}}:
\batitle{Perspective on superconducting qubit quantum computing}.
\bjtitle{The European Physical Journal A}
\bvolume{59}(\bissue{5}),
\bfpage{94}
(\byear{2023})
\end{barticle}
\endbibitem

\bibitem[\protect\citeauthoryear{Sussman}{2023}]{sussman2023qc}
\begin{botherref}
\oauthor{\bsnm{Sussman}, \binits{S.}}:
Quantum computing with an open source qubit controller.
PhD thesis,
Princeton University
(2023)
\end{botherref}
\endbibitem

\bibitem[\protect\citeauthoryear{McKay et~al.}{2023}]{mckay2023}
\begin{botherref}
\oauthor{\bsnm{McKay}, \binits{D.C.}},
\oauthor{\bsnm{Hincks}, \binits{I.}},
\oauthor{\bsnm{Pritchett}, \binits{E.J.}},
\oauthor{\bsnm{Carroll}, \binits{M.}},
\oauthor{\bsnm{Govia}, \binits{L.C.G.}},
\oauthor{\bsnm{Merkel}, \binits{S.T.}}:
Benchmarking Quantum Processor Performance at Scale
(2023)
\end{botherref}
\endbibitem

\bibitem[\protect\citeauthoryear{Johansson et~al.}{2021}]{johansson2021qc}
\begin{botherref}
\oauthor{\bsnm{Johansson}, \binits{M.P.}},
\oauthor{\bsnm{Krishnasamy}, \binits{E.}},
\oauthor{\bsnm{Meyer}, \binits{N.}},
\oauthor{\bsnm{Piechurski}, \binits{C.}}:
Quantum computing--a european perspective.
PRACE-6IP TR
(2021)
\end{botherref}
\endbibitem

\bibitem[\protect\citeauthoryear{Jyothi~Ahuja and Dutt}{2022}]{jyothi2022io}
\begin{botherref}
\oauthor{\bsnm{Jyothi~Ahuja}, \binits{N.}},
\oauthor{\bsnm{Dutt}, \binits{S.}}:
Implications of quantum science on industry 4.0: Challenges and opportunities.
Quantum and Blockchain for Modern Computing Systems: Vision and Advancements: Quantum and Blockchain Technologies: Current Trends and Challenges,
183--204
(2022)
\end{botherref}
\endbibitem

\bibitem[\protect\citeauthoryear{Qin et~al.}{2022}]{qin2022ao}
\begin{botherref}
\oauthor{\bsnm{Qin}, \binits{D.}},
\oauthor{\bsnm{Xu}, \binits{X.}},
\oauthor{\bsnm{Li}, \binits{Y.}}:
An overview of quantum error mitigation formulas.
Chinese Physics B
(2022)
\end{botherref}
\endbibitem

\bibitem[\protect\citeauthoryear{Van Den~Berg et~al.}{2023}]{van2023pe}
\begin{botherref}
\oauthor{\bsnm{Van Den~Berg}, \binits{E.}},
\oauthor{\bsnm{Minev}, \binits{Z.K.}},
\oauthor{\bsnm{Kandala}, \binits{A.}},
\oauthor{\bsnm{Temme}, \binits{K.}}:
Probabilistic error cancellation with sparse pauli--lindblad models on noisy quantum processors.
Nature Physics,
1--6
(2023)
\end{botherref}
\endbibitem

\bibitem[\protect\citeauthoryear{Temme et~al.}{2017}]{temme2017em}
\begin{barticle}
\bauthor{\bsnm{Temme}, \binits{K.}},
\bauthor{\bsnm{Bravyi}, \binits{S.}},
\bauthor{\bsnm{Gambetta}, \binits{J.M.}}:
\batitle{Error mitigation for short-depth quantum circuits}.
\bjtitle{Physical review letters}
\bvolume{119}(\bissue{18}),
\bfpage{180509}
(\byear{2017})
\end{barticle}
\endbibitem

\bibitem[\protect\citeauthoryear{Ramadhani et~al.}{2021}]{ramadhani2021qe}
\begin{barticle}
\bauthor{\bsnm{Ramadhani}, \binits{S.}},
\bauthor{\bsnm{Rehman}, \binits{J.U.}},
\bauthor{\bsnm{Shin}, \binits{H.}}:
\batitle{Quantum error mitigation for quantum state tomography}.
\bjtitle{IEEE Access}
\bvolume{9},
\bfpage{107955}--\blpage{107964}
(\byear{2021})
\end{barticle}
\endbibitem

\bibitem[\protect\citeauthoryear{Tannu and Qureshi}{2019}]{tannu2019mm}
\begin{bchapter}
\bauthor{\bsnm{Tannu}, \binits{S.S.}},
\bauthor{\bsnm{Qureshi}, \binits{M.K.}}:
\bctitle{Mitigating measurement errors in quantum computers by exploiting state-dependent bias}.
In: \bbtitle{Proceedings of the 52nd Annual IEEE/ACM International Symposium on Microarchitecture},
pp. \bfpage{279}--\blpage{290}
(\byear{2019})
\end{bchapter}
\endbibitem

\bibitem[\protect\citeauthoryear{Sun et~al.}{2021}]{sun2021mr}
\begin{barticle}
\bauthor{\bsnm{Sun}, \binits{J.}},
\bauthor{\bsnm{Yuan}, \binits{X.}},
\bauthor{\bsnm{Tsunoda}, \binits{T.}},
\bauthor{\bsnm{Vedral}, \binits{V.}},
\bauthor{\bsnm{Benjamin}, \binits{S.C.}},
\bauthor{\bsnm{Endo}, \binits{S.}}:
\batitle{Mitigating realistic noise in practical noisy intermediate-scale quantum devices}.
\bjtitle{Physical Review Applied}
\bvolume{15}(\bissue{3}),
\bfpage{034026}
(\byear{2021})
\end{barticle}
\endbibitem

\bibitem[\protect\citeauthoryear{Niu and Todri-Sanial}{2022}]{niu2022eo}
\begin{barticle}
\bauthor{\bsnm{Niu}, \binits{S.}},
\bauthor{\bsnm{Todri-Sanial}, \binits{A.}}:
\batitle{Effects of dynamical decoupling and pulse-level optimizations on ibm quantum computers}.
\bjtitle{IEEE Transactions on Quantum Engineering}
\bvolume{3},
\bfpage{1}--\blpage{10}
(\byear{2022})
\end{barticle}
\endbibitem

\bibitem[\protect\citeauthoryear{Qi et~al.}{2023}]{qi2023eo}
\begin{barticle}
\bauthor{\bsnm{Qi}, \binits{J.}},
\bauthor{\bsnm{Xu}, \binits{X.}},
\bauthor{\bsnm{Poletti}, \binits{D.}},
\bauthor{\bsnm{Ng}, \binits{H.K.}}:
\batitle{Efficacy of noisy dynamical decoupling}.
\bjtitle{Physical Review A}
\bvolume{107}(\bissue{3}),
\bfpage{032615}
(\byear{2023})
\end{barticle}
\endbibitem

\bibitem[\protect\citeauthoryear{Hartnett et~al.}{2023}]{hartnett2023is}
\begin{botherref}
\oauthor{\bsnm{Hartnett}, \binits{G.}},
\oauthor{\bsnm{Mundada}, \binits{P.}},
\oauthor{\bsnm{Baum}, \binits{Y.}},
\oauthor{\bsnm{Kakkar}, \binits{A.}},
\oauthor{\bsnm{Stace}, \binits{T.}}:
Improving syndrome detection using quantum optimal control.
Bulletin of the American Physical Society
(2023)
\end{botherref}
\endbibitem

\bibitem[\protect\citeauthoryear{Dong et~al.}{2020}]{dong2020rc}
\begin{barticle}
\bauthor{\bsnm{Dong}, \binits{Y.}},
\bauthor{\bsnm{Meng}, \binits{X.}},
\bauthor{\bsnm{Lin}, \binits{L.}},
\bauthor{\bsnm{Kosut}, \binits{R.}},
\bauthor{\bsnm{Whaley}, \binits{K.B.}}:
\batitle{Robust control optimization for quantum approximate optimization algorithms}.
\bjtitle{IFAC-PapersOnLine}
\bvolume{53}(\bissue{2}),
\bfpage{242}--\blpage{249}
(\byear{2020})
\end{barticle}
\endbibitem

\bibitem[\protect\citeauthoryear{Hashim et~al.}{2020}]{hashim2020rc}
\begin{botherref}
\oauthor{\bsnm{Hashim}, \binits{A.}},
\oauthor{\bsnm{Naik}, \binits{R.K.}},
\oauthor{\bsnm{Morvan}, \binits{A.}},
\oauthor{\bsnm{Ville}, \binits{J.-L.}},
\oauthor{\bsnm{Mitchell}, \binits{B.}},
\oauthor{\bsnm{Kreikebaum}, \binits{J.M.}},
\oauthor{\bsnm{Davis}, \binits{M.}},
\oauthor{\bsnm{Smith}, \binits{E.}},
\oauthor{\bsnm{Iancu}, \binits{C.}},
\oauthor{\bsnm{O'Brien}, \binits{K.P.}}, et al.:
Randomized compiling for scalable quantum computing on a noisy superconducting quantum processor.
arXiv preprint arXiv:2010.00215
(2020)
\end{botherref}
\endbibitem

\bibitem[\protect\citeauthoryear{Urbanek et~al.}{2021}]{urbanek2021md}
\begin{barticle}
\bauthor{\bsnm{Urbanek}, \binits{M.}},
\bauthor{\bsnm{Nachman}, \binits{B.}},
\bauthor{\bsnm{Pascuzzi}, \binits{V.R.}},
\bauthor{\bsnm{He}, \binits{A.}},
\bauthor{\bsnm{Bauer}, \binits{C.W.}},
\bauthor{\bsnm{Jong}, \binits{W.A.}}:
\batitle{Mitigating depolarizing noise on quantum computers with noise-estimation circuits}.
\bjtitle{Physical review letters}
\bvolume{127}(\bissue{27}),
\bfpage{270502}
(\byear{2021})
\end{barticle}
\endbibitem

\bibitem[\protect\citeauthoryear{Ware et~al.}{2021}]{ware2021ep}
\begin{barticle}
\bauthor{\bsnm{Ware}, \binits{M.}},
\bauthor{\bsnm{Ribeill}, \binits{G.}},
\bauthor{\bsnm{Riste}, \binits{D.}},
\bauthor{\bsnm{Ryan}, \binits{C.A.}},
\bauthor{\bsnm{Johnson}, \binits{B.}},
\bauthor{\bsnm{Da~Silva}, \binits{M.P.}}:
\batitle{Experimental pauli-frame randomization on a superconducting qubit}.
\bjtitle{Physical Review A}
\bvolume{103}(\bissue{4}),
\bfpage{042604}
(\byear{2021})
\end{barticle}
\endbibitem

\bibitem[\protect\citeauthoryear{Suzuki et~al.}{2022}]{suzuki2022qe}
\begin{barticle}
\bauthor{\bsnm{Suzuki}, \binits{Y.}},
\bauthor{\bsnm{Endo}, \binits{S.}},
\bauthor{\bsnm{Fujii}, \binits{K.}},
\bauthor{\bsnm{Tokunaga}, \binits{Y.}}:
\batitle{Quantum error mitigation as a universal error reduction technique: Applications from the nisq to the fault-tolerant quantum computing eras}.
\bjtitle{PRX Quantum}
\bvolume{3}(\bissue{1}),
\bfpage{010345}
(\byear{2022})
\end{barticle}
\endbibitem

\bibitem[\protect\citeauthoryear{Endo et~al.}{2018}]{endo2018pq}
\begin{barticle}
\bauthor{\bsnm{Endo}, \binits{S.}},
\bauthor{\bsnm{Benjamin}, \binits{S.C.}},
\bauthor{\bsnm{Li}, \binits{Y.}}:
\batitle{Practical quantum error mitigation for near-future applications}.
\bjtitle{Physical Review X}
\bvolume{8}(\bissue{3}),
\bfpage{031027}
(\byear{2018})
\end{barticle}
\endbibitem

\bibitem[\protect\citeauthoryear{Cai et~al.}{2023}]{cai2023qe}
\begin{barticle}
\bauthor{\bsnm{Cai}, \binits{Z.}},
\bauthor{\bsnm{Babbush}, \binits{R.}},
\bauthor{\bsnm{Benjamin}, \binits{S.C.}},
\bauthor{\bsnm{Endo}, \binits{S.}},
\bauthor{\bsnm{Huggins}, \binits{W.J.}},
\bauthor{\bsnm{Li}, \binits{Y.}},
\bauthor{\bsnm{McClean}, \binits{J.R.}},
\bauthor{\bsnm{O’Brien}, \binits{T.E.}}:
\batitle{Quantum error mitigation}.
\bjtitle{Reviews of Modern Physics}
\bvolume{95}(\bissue{4}),
\bfpage{045005}
(\byear{2023})
\end{barticle}
\endbibitem

\bibitem[\protect\citeauthoryear{Strikis et~al.}{2021}]{strikis2021lb}
\begin{barticle}
\bauthor{\bsnm{Strikis}, \binits{A.}},
\bauthor{\bsnm{Qin}, \binits{D.}},
\bauthor{\bsnm{Chen}, \binits{Y.}},
\bauthor{\bsnm{Benjamin}, \binits{S.C.}},
\bauthor{\bsnm{Li}, \binits{Y.}}:
\batitle{Learning-based quantum error mitigation}.
\bjtitle{PRX Quantum}
\bvolume{2}(\bissue{4}),
\bfpage{040330}
(\byear{2021})
\end{barticle}
\endbibitem

\bibitem[\protect\citeauthoryear{Endo et~al.}{2021}]{endo2021hq}
\begin{barticle}
\bauthor{\bsnm{Endo}, \binits{S.}},
\bauthor{\bsnm{Cai}, \binits{Z.}},
\bauthor{\bsnm{Benjamin}, \binits{S.C.}},
\bauthor{\bsnm{Yuan}, \binits{X.}}:
\batitle{Hybrid quantum-classical algorithms and quantum error mitigation}.
\bjtitle{Journal of the Physical Society of Japan}
\bvolume{90}(\bissue{3}),
\bfpage{032001}
(\byear{2021})
\end{barticle}
\endbibitem

\bibitem[\protect\citeauthoryear{Takagi et~al.}{2022}]{takagi2022fl}
\begin{barticle}
\bauthor{\bsnm{Takagi}, \binits{R.}},
\bauthor{\bsnm{Endo}, \binits{S.}},
\bauthor{\bsnm{Minagawa}, \binits{S.}},
\bauthor{\bsnm{Gu}, \binits{M.}}:
\batitle{Fundamental limits of quantum error mitigation}.
\bjtitle{npj Quantum Information}
\bvolume{8}(\bissue{1}),
\bfpage{114}
(\byear{2022})
\end{barticle}
\endbibitem

\bibitem[\protect\citeauthoryear{Chen}{2023}]{chen2023sd}
\begin{bchapter}
\bauthor{\bsnm{Chen}, \binits{K.-C.}}:
\bctitle{Short-depth circuits and error mitigation for large-scale ghz-state preparation, and benchmarking on ibm's 127-qubit system}.
In: \bbtitle{2023 IEEE International Conference on Quantum Computing and Engineering (QCE)},
vol. \bseriesno{2},
pp. \bfpage{207}--\blpage{210}
(\byear{2023}).
\bcomment{IEEE}
\end{bchapter}
\endbibitem

\bibitem[\protect\citeauthoryear{Bruzewicz et~al.}{2019}]{bruzewicz2019ti}
\begin{botherref}
\oauthor{\bsnm{Bruzewicz}, \binits{C.D.}},
\oauthor{\bsnm{Chiaverini}, \binits{J.}},
\oauthor{\bsnm{McConnell}, \binits{R.}},
\oauthor{\bsnm{Sage}, \binits{J.M.}}:
Trapped-ion quantum computing: Progress and challenges.
Applied Physics Reviews
\textbf{6}(2)
(2019)
\end{botherref}
\endbibitem

\bibitem[\protect\citeauthoryear{Chen et~al.}{2022}]{pulse_zne}
\begin{barticle}
\bauthor{\bsnm{Chen}, \binits{I.-C.}},
\bauthor{\bsnm{Burdick}, \binits{B.}},
\bauthor{\bsnm{Yao}, \binits{Y.}},
\bauthor{\bsnm{Orth}, \binits{P.P.}},
\bauthor{\bsnm{Iadecola}, \binits{T.}}:
\batitle{Error-mitigated simulation of quantum many-body scars on quantum computers with pulse-level control}.
\bjtitle{Phys. Rev. Res.}
\bvolume{4},
\bfpage{043027}
(\byear{2022})
\doiurl{10.1103/PhysRevResearch.4.043027}
\end{barticle}
\endbibitem

\bibitem[\protect\citeauthoryear{Giurgica-Tiron et~al.}{2020}]{giurgica2020dz}
\begin{bchapter}
\bauthor{\bsnm{Giurgica-Tiron}, \binits{T.}},
\bauthor{\bsnm{Hindy}, \binits{Y.}},
\bauthor{\bsnm{LaRose}, \binits{R.}},
\bauthor{\bsnm{Mari}, \binits{A.}},
\bauthor{\bsnm{Zeng}, \binits{W.J.}}:
\bctitle{Digital zero noise extrapolation for quantum error mitigation}.
In: \bbtitle{2020 IEEE International Conference on Quantum Computing and Engineering (QCE)},
pp. \bfpage{306}--\blpage{316}
(\byear{2020}).
\bcomment{IEEE}
\end{bchapter}
\endbibitem

\bibitem[\protect\citeauthoryear{Pascuzzi et~al.}{2022}]{pascuzzi2022ce}
\begin{barticle}
\bauthor{\bsnm{Pascuzzi}, \binits{V.R.}},
\bauthor{\bsnm{He}, \binits{A.}},
\bauthor{\bsnm{Bauer}, \binits{C.W.}},
\bauthor{\bsnm{De~Jong}, \binits{W.A.}},
\bauthor{\bsnm{Nachman}, \binits{B.}}:
\batitle{Computationally efficient zero-noise extrapolation for quantum-gate-error mitigation}.
\bjtitle{Physical Review A}
\bvolume{105}(\bissue{4}),
\bfpage{042406}
(\byear{2022})
\end{barticle}
\endbibitem

\bibitem[\protect\citeauthoryear{He et~al.}{2020}]{he2020zn}
\begin{barticle}
\bauthor{\bsnm{He}, \binits{A.}},
\bauthor{\bsnm{Nachman}, \binits{B.}},
\bauthor{\bsnm{Jong}, \binits{W.A.}},
\bauthor{\bsnm{Bauer}, \binits{C.W.}}:
\batitle{Zero-noise extrapolation for quantum-gate error mitigation with identity insertions}.
\bjtitle{Physical Review A}
\bvolume{102}(\bissue{1}),
\bfpage{012426}
(\byear{2020})
\end{barticle}
\endbibitem

\bibitem[\protect\citeauthoryear{Murali et~al.}{2019}]{noise_adaptive}
\begin{bchapter}
\bauthor{\bsnm{Murali}, \binits{P.}},
\bauthor{\bsnm{Baker}, \binits{J.M.}},
\bauthor{\bsnm{Javadi-Abhari}, \binits{A.}},
\bauthor{\bsnm{Chong}, \binits{F.T.}},
\bauthor{\bsnm{Martonosi}, \binits{M.}}:
\bctitle{Noise-adaptive compiler mappings for noisy intermediate-scale quantum computers}.
In: \bbtitle{Proceedings of the Twenty-Fourth International Conference on Architectural Support for Programming Languages and Operating Systems}.
\bsertitle{ASPLOS '19},
pp. \bfpage{1015}--\blpage{1029}.
\bpublisher{Association for Computing Machinery},
\blocation{New York, NY, USA}
(\byear{2019}).
\doiurl{10.1145/3297858.3304075} .
\burl{https://doi.org/10.1145/3297858.3304075}
\end{bchapter}
\endbibitem

\bibitem[\protect\citeauthoryear{{Qiskit contributors}}{2023}]{Qiskit}
\begin{botherref}
\oauthor{\bsnm{{Qiskit contributors}}}:
Qiskit: An Open-source Framework for Quantum Computing
(2023).
\doiurl{10.5281/zenodo.2573505}
\end{botherref}
\endbibitem

\bibitem[\protect\citeauthoryear{Bernstein and Vazirani}{1997}]{bernstein}
\begin{barticle}
\bauthor{\bsnm{Bernstein}, \binits{E.}},
\bauthor{\bsnm{Vazirani}, \binits{U.}}:
\batitle{Quantum complexity theory}.
\bjtitle{SIAM Journal on Computing}
\bvolume{26}(\bissue{5}),
\bfpage{1411}--\blpage{1473}
(\byear{1997})
\doiurl{10.1137/S0097539796300921}
{\href{https://arxiv.org/abs/https://doi.org/10.1137/S0097539796300921}{{https://doi.org/10.1137/S0097539796300921}}}
\end{barticle}
\endbibitem

\bibitem[\protect\citeauthoryear{LaRose et~al.}{2022}]{LaRose_2022}
\begin{barticle}
\bauthor{\bsnm{LaRose}, \binits{R.}},
\bauthor{\bsnm{Mari}, \binits{A.}},
\bauthor{\bsnm{Kaiser}, \binits{S.}},
\bauthor{\bsnm{Karalekas}, \binits{P.J.}},
\bauthor{\bsnm{Alves}, \binits{A.A.}},
\bauthor{\bsnm{Czarnik}, \binits{P.}},
\bauthor{\bsnm{El~Mandouh}, \binits{M.}},
\bauthor{\bsnm{Gordon}, \binits{M.H.}},
\bauthor{\bsnm{Hindy}, \binits{Y.}},
\bauthor{\bsnm{Robertson}, \binits{A.}},
\bauthor{\bsnm{Thakre}, \binits{P.}},
\bauthor{\bsnm{Wahl}, \binits{M.}},
\bauthor{\bsnm{Samuel}, \binits{D.}},
\bauthor{\bsnm{Mistri}, \binits{R.}},
\bauthor{\bsnm{Tremblay}, \binits{M.}},
\bauthor{\bsnm{Gardner}, \binits{N.}},
\bauthor{\bsnm{Stemen}, \binits{N.T.}},
\bauthor{\bsnm{Shammah}, \binits{N.}},
\bauthor{\bsnm{Zeng}, \binits{W.J.}}:
\batitle{Mitiq: A software package for error mitigation on noisy quantum computers}.
\bjtitle{Quantum}
\bvolume{6},
\bfpage{774}
(\byear{2022})
\doiurl{10.22331/q-2022-08-11-774}
\end{barticle}
\endbibitem

\end{thebibliography}

\end{document}